\documentstyle[12pt]{article}
\newcommand{\si}{{\sigma}}
\newcommand{\om}{{\omega}}

\newcommand{\bfq}{\mbox{\boldmath {$q$}}}
\newcommand{\bfp}{\mbox{\boldmath {$p$}}}

\newcommand{\fslash}[1]{{\mbox{$\!\not\!#1$}}}

\newcommand{\cE}{{\cal E}}

\newcommand{\mPi}{{\mit \Pi}}

\newcommand{\mSigma}{{\mit \Sigma}}

\newcommand{\rmT}{{\rm T}}
\newcommand{\rmF}{{\rm F}}
\newcommand{\rmD}{{\rm D}}

\newcommand{\rmI}{{\rm I}}

\newcommand{\rmP}{{\rm P}}

\newcommand{\rmL}{{\rm L}}
\newcommand{\rmM}{{\rm M}}

\newcommand{\rmW}{{\rm W}}

\newcommand{\rmc}{{\rm c}}

\newcommand{\rmd}{{\rm d}}
\newcommand{\rme}{{\rm e}}

\newcommand{\rmx}{{\rm x}}
\newcommand{\rmsc}{{\rm sc}}

\newcommand{\gaq}{\stackrel{\displaystyle{>}}{\sim}}
\newcommand{\refer}[1]{(\ref{#1})}

\newcommand{\Msta}{{M^{\ast}}}

\newcommand{\pF}{p_{\rmF}}
\newcommand{\vF}{v_{\rmF}}

\newcommand{\ro}{\varrho}

\begin{document}
\title{Quasiparticle properties and the dynamics of
       high-density nuclear matter}
\author{Kazuhiro TANAKA\thanks{Special Researcher, Basic Science Program}
          \\
        {\mbox{$ $}}\\
        {\sl Radiation Laboratory}\\
        {\sl The Institute of Physical and Chemical
             Research (RIKEN)}\\
        {\sl Hirosawa 2-1, Wako-shi, Saitama, 351-01 Japan}}
\date{ }
\maketitle
\begin{abstract}
The energy spectrum of nucleons in high-density nuclear matter
is investigated in the framework of relativistic meson-nucleon
many-body theory, employing the $1/N$ expansion method.
The coupling of the nucleon with the particle-hole
excitations in the medium flattens
the spectra in the vicinity of the Fermi surface.
The effect
grows logarithmically for increasing density
and eventually leads to instability of the normal state.
The validity of the mean-field theory
at high density is criticized.
\end{abstract}

\baselineskip 3.7ex

\newpage
\setcounter{equation}{0}

\indent
One universal character shared by most normal
Fermi liquids is the enhancement of the effective mass
around the Fermi surface \cite{ma}.
Typical indications observed empirically are
the density of single-particle levels around the ``Fermi
surface'' in nuclei, and the logarithmic ($T^{3}\ln T$)
correction to the specific heat at low temperature $T$
in normal liquid $^{3}$He.
It is now well established that these phenomena
due to the energy-dependence of the effective mass
reflect the dynamics beyond the mean-field theory:
The relevant mechanism is the coupling of the single-particle
motion to other excitation modes, e.g.,
particle-hole (ph) excitations in the case of nuclear matter.
As a result,
the nucleon effective mass at normal density
is close to the free mass $M$ or slightly larger
at the Fermi surface
while it is about the traditional value ($\sim 0.7 M$)
away from the Fermi surface.

The effective mass at the Fermi surface determines
the density of states at the
Fermi surface. Therefore, the dynamical couplings to
the single-particle motion could influence the observables
significantly. In this light, the density-dependence,
especially the high-density behavior,
of the effective mass would be relevant
to descriptions of high-density matter,
needed
to understand, e.g., stellar structure or evolution and
high-energy heavy ion reactions.

The investigation of the high-density behavior of the effective
mass requires a relativistic framework of Fermi liquids.
In this paper
we compute the quasiparticle spectra
including higher-order many-body correlations
employing a relativistic many-body theory:
In order to exhibit clearly the points avoiding complicated
aspects such as the compositeness of hadrons and
the possibilities of new degrees of freedom at high density,
we use the simplest relativistic model of nuclear matter,
the ``$\si \om$ model'' \cite{se}:
\begin{equation}
{\cal L} = \overline{\psi} \left( i \fslash{\partial}
- g_{\om} \fslash{\om} - M + g_{\si} \si \right) \psi
+ \frac{1}{2} \left( \partial_{\mu} \si \right)^{2}
- \frac{1}{2} m_{\si}^{2} \si^{2} - \frac{1}{4}\om_{\mu \nu}^{2}
+ \frac{1}{2} m_{\om}^{2} \om_{\mu}^{2},
\label{eq:la}
\end{equation}
with $\om_{\mu \nu} = \partial_{\mu}\om_{\nu} - \partial_{\nu} \om_{\mu}$.
$\psi$, $\si$ and $\om_{\mu}$ are the nucleon, $\si$- and $\om$-meson
fields.

Since this model is renormalizable, it
allows a systematic investigation of the many-body
correlations.
One convenient scheme for this purpose
is the $1/N$ expansion where $N$ is
the extended number of nucleon species \cite{ta3}.
The expansion
can be obtained
by applying
the rescaling: $g_{\si} \rightarrow 1/\sqrt{N} g_{\si}$,
$g_{\om} \rightarrow 1/\sqrt{N} g_{\om}$;
$\si \rightarrow \sqrt{N} \si$,
$\om_{\mu} \rightarrow \sqrt{N} \om_{\mu}$
in the lagrangian \refer{eq:la}.
The energy density of nuclear matter
up to the next-to-leading order
is given by
$\cE = \cE_{0} + \cE_{\rm e} + \cE_{\rm c}$
\cite{ta3,ta4}.
$\cE_{0}$ denotes the Hartree term which is the
leading order of $O(N)$,
while $\cE_{\rm e}$ and $\cE_{\rm c}$ denote the
exchange and the RPA-type correlation energy terms, both of which
are of $O(1)$.
The formulae for these terms
can be found in ref. \cite{ta3}.
Based on this framework,
a successful description of the saturation property,
the equation of state,
and the Fermi-liquid properties of nuclear matter
has been obtained \cite{ta3,ta4,he},
improving the Hartree description.
The essential ingredients
are the many-body correlation effects
due to
$\cE_{\rm c}$.

The quasiparticle energy corresponding to our energy density
can be obtained by taking the functional derivative
with respect to the Fermi distribution function $n(\bfp)$
following the Landau theory of Fermi liquids \cite{la1}
and its relativistic extension \cite{ba3}:
\begin{equation}
\epsilon (\bfp) = \frac{\delta \cE}{\delta n(\bfp)}
= \epsilon_{0} (\bfp) + \epsilon_{\rm e}(\bfp)
+ \epsilon_{\rm c}(\bfp),
\label{eq:qp2}
\end{equation}
where the three terms derive from the corresponding three terms
of our energy density.
The familiar Hartree term $\epsilon_{0}$, which is of $O(1)$,
contains the scalar and
the vector mean-field potential contributions.
The other contributions of $O(1/N)$ are given by
\cite{ta4}
\begin{equation}
\epsilon_{\rm x}(\bfp) = \frac{\Msta}{E^{\ast}(p)}
\overline{u}(p) \mSigma_{\rm x}(p) u(p)
\;\;\; \;\;\; ({\rm x} = {\rm e}, \; {\rm c}),
\label{eq:e1}
\end{equation}
where $E^{\ast}(p) = \sqrt{\bfp^{2} + {M^{\ast}}^{2}}$
and $\Msta = M - g_{\si} \overline{\si}$ with $\overline{\si}$
the ground-state expectation value of the $\si$-field.
$u(p)$ is the Hartree spinor
normalized by $\overline{u}u = 1$.
$\mSigma_{\rm x}$ is the nucleon self-energy
shown in fig. \ref{fig:fig2}; ${\rm x} = {\rm e}$ and
${\rm c}$ correspond to the first (exchange) and
the following (RPA) graphs,
respectively.\footnote{For convenience of the following discussion,
we employ a slightly different grouping of
the next-to-leading terms
compared to the preceding works
\cite{ta3,ta4}.}

The counter terms to renormalize
the divergences of the self-energies
are determined through eq. \refer{eq:qp2}
consistently with the renormalization conditions for the energy density:
We renormalize all vertex parts in free space
at the scales $q^{2} = 0$ and $\fslash{p} = M$ for the meson
and the nucleon legs, respectively.
The Landau ghost singularity
appearing in the one-loop meson propagators
is removed by the method based on the
K\"{a}ll\'{e}n-Lehmann representation
proposed by Redmond \cite{re}
and extended to finite density in refs. \cite{ta1,ta3,ta4}.

If the nucleon Hartree propagator connecting
the external legs in the diagrams of fig. \ref{fig:fig2} is separated
into the ``Feynman part'' (F)
and the ``density-dependent part'' (D) \cite{se},
we obtain the conventional separation of $\mSigma_{\rmx}$
into $\mSigma_{\rmx \rmF}$
and $\mSigma_{\rmx \rmD}$.
For the present purpose,
it is convenient to employ a new separation of the self-energy
based on a Wick rotation (WR) of the loop integration
in $\mSigma_{\rmx \rmF}$ \cite{ta4,bl}:
$\epsilon_{\rmx}(\bfp) = \epsilon_{\rmx \rmW}(\bfp)
+ \epsilon_{\rmx \rmP}(\bfp)$, where
$\epsilon_{\rmx \rmW}$ involves the Wick rotated
euclidian 4-integral, and
$\epsilon_{\rmx \rmP}$ contains an on-shell intermediate nucleon line,
arising from the nucleon pole
due to the WR and the contribution of $\mSigma_{\rmx \rmD}$.
$\epsilon_{\rmx \rmP}$ can be expressed compactly:
\begin{equation}
\epsilon_{\rmx \rmP} (\bfp)= 4 \int^{\pF}_{p}
\frac{\rmd q q^{2}}{(2 \pi)^{2}} \int^{1}_{-1}
\rmd \! \cos \theta \: f_{\rmx}(\bfp, \bfq),
\label{eq:po}
\end{equation}
with $\theta$ the angle between $\bfp$ and $\bfq$.
$f_{\rmx}(\bfp, \bfq)$ are the exchange-type interactions
between two quasiparticles with momenta
$\bfp$, $\bfq$.
$f_{\rm e}$ involves the noninteracting meson propagators,
while $f_{\rm c}$ takes into account the modifications due to the
one-loop meson self-energies.

The slope of the quasiparticle spectra at the Fermi surface
defines the Fermi velocity:
\begin{equation}
\vF = \left. \frac{\partial \epsilon(\bfp)}{\partial |\bfp|}
\right|_{|\bfp| = \pF}
= \frac{\pF}{E^{\ast}(\pF)} + v_{\rmF \rm e} + v_{\rmF \rm c},
\label{eq:vf}
\end{equation}
where the three terms derive from the corresponding terms of
eq. \refer{eq:qp2}.
In the NR treatment, $\vF \equiv \pF/M_{\rm eff}$
with $M_{\rm eff}$ the effective mass at the
Fermi surface.
If, as we will show below,
the enhancement of $M_{\rm eff}$ becomes stronger and stronger
with increasing density,
$v_{\rm F}$ can eventually become a decreasing function
of the density. In particular, $v_{\rm F} \rightarrow 0$
means that the spectra become flat around the Fermi surface.

In fig. \ref{fig:fig4} we show $\vF$ as a
function of $\pF$. The parameters are taken from refs. \cite{ta3,ta4}
and are shown in table \ref{tab:tabn}:
The dotted line shows the Hartree approximation, i.e.,
the first term of eq. \refer{eq:vf} using the ``Hartree'' parameter set;
it approaches the causal limit
``1''
at high density.
In contrast, the solid line, which shows the full result
of \refer{eq:vf} using the parameter set B,
crosses $\vF = 0$ at the critical
$\pF = p_{\rmF}^{\rmc}$, and $\vF$ becomes negative
for higher densities.
The dashed line, including only one density-dependent bubble
instead of the full RPA
propagator,\footnote{The Feynman bubbles are summed up in this result.
}
shows a
behavior similar to the full case, but $p_{\rmF}^{\rmc}$
becomes smaller.
We also show the full result of \refer{eq:vf} using the parameter
set A by the dot-dashed line; the comparison with the solid line
demonstrates the sensitivity of $p_{\rmF}^{\rmc}$ to the
parameters (see eq. \refer{eq:d3} below).
In fig. \ref{fig:fig5} we show the
quasiparticle energy of eq. \refer{eq:qp2}
for the case of the solid line of fig. \ref{fig:fig4}.
For high densities, the spectra
show the ``anomaly'' around the Fermi surface, i.e.,
they are flat or decreasing
due to the inclusion of the higher-order self-energy
terms.\footnote{The latter case with
the negative $\vF$ does not correspond to the normal state
and therefore, strictly speaking, the Landau theory is not applicable.
This point will be discussed later.}
Thus, the effect
grows with increasing density.

{}From fig. \ref{fig:fig5}(b),
we see that this effect is due to $\epsilon_{\rmc \rmP}$,
which is a decreasing function around the Fermi surface.
(This is seen
already at normal density,
but
for higher
densities this effect becomes more pronounced
and shows up in the total results.)
The decrease of $\epsilon_{\rmc \rmP}$
around the Fermi surface could be expected from
eq. \refer{eq:po} if the result of the
angular average of $f_{\rmx} (\bfp, \bfq)$
assumes a positive value (see also ref. \cite{bl} for the NR case).

Now can we understand the origin of the growth of the anomaly
due to $\epsilon_{\rmc \rmP}$?
For this purpose it will be useful to examine
the leading high-density behavior of $\vF$.
We set for $\pF \rightarrow \infty$
$\vF \rightarrow 1 + \delta v_{\rmF \rme}
+ \delta v_{\rmF \rmc} \equiv 1 + \delta v_{\rmF}$,
where ``1'' is due to the first (Hartree) term of eq. \refer{eq:vf}
while $\delta v_{\rmF \rmx}$ ($\rmx = \rme$, $\rmc$)
denote the asymptotic limits of the following
two ($O(1/N)$) terms $v_{\rmF \rme}$, $v_{\rmF \rmc}$:
\begin{equation}
\delta v_{\rmF \rmx} \approx
- \frac{p_{\rmF}^{2}}{\pi^{2}}
\int_{-1}^{1} \rmd \! \cos \theta \: \left.
f_{\rmx} (\bfp, \bfq)
\right|_{p = q = \pF}.
\label{eq:vfi}
\end{equation}
The r.h.s. is the result
obtained by substituting $\epsilon_{\rmx \rmP}$
of eq. \refer{eq:po} into eq. \refer{eq:vf}.
``$\approx$'' means that
the contribution due to $\epsilon_{\rmx \rmW}$ is neglected
because of the above discussion of fig. \ref{fig:fig5}.
(In fact, it can be shown
analytically that $\epsilon_{\rmx \rmW}$ does not alter our conclusions
\cite{ta5}.)

It is convenient to separate the integrand of eq. \refer{eq:vfi}
according to the types of the possible
meson modes in the medium \cite{ta4}:
\begin{equation}
\left. f_{\rmx}(\bfp, \bfq) \right|_{p = q = \pF}
= - \frac{1}{8 p_{\rmF}^{2}}
\sum_{\rmI} A_{\rmx \rmI}(\cos \theta)
h_{\rmI}(\cos \theta),
\label{eq:fs}
\end{equation}
where $\rmI = \si, \rmM, \rmL$ and $\rmT$,
corresponding to the $\si$-, mixed, longitudinal
and transverse contributions.
$h_{\rmI}$ is a dimensionless function
accounting for
the scalar and the vector vertex structures.
$A_{\rme \rmI}$
are the exchange contributions due to the
noninteracting
$\si$- and $\om$-meson propagators
while $A_{\rmc \rmI}$ give the modifications of $A_{\rme \rmI}$
due to the RPA correlations.

To extract the leading asymptotic behavior
of \refer{eq:fs},
we set $\Msta = 0$; because there is no singularity
for $\Msta = 0$
and from the dimension counting,
any contributions omitted here will be suppressed for
$\pF \rightarrow \infty$ compared to the retained ones.
In this limit it can be easily seen that the contributions
due to the M-mode are suppressed compared to the others.

In this case, the relevant contributions
($\rmI = \si$, $\rmL$ and $\rmT$) have the form:
\begin{equation}
a^{n}_{\rmI} = \frac{g_{\rmI}^{2}}{2}
\frac{1}{t + \kappa_{\rmI}^{2}}
\left(\frac{\hat{\mPi}_{\rmI}(t)}{t + \kappa_{\rmI}^{2}}
\right)^{n},
\label{eq:yu}
\end{equation}
where
$t = 1- \cos \theta$,
$g_{\rmL} = g_{\rmT} \equiv g_{\om}$,
$m_{\rmL}= m_{\rmT} \equiv m_{\om}$,
$\kappa_{\rmI}^{2} = m_{\rmI}^{2}/2 p_{\rmF}^{2}$,
and $\hat{\mPi}_{\rmI}(t) = \left. \mPi_{\rmI}(p-q)
\right|_{p = q = \pF}/2 p_{\rmF}^{2}$ with
$\mPi_{\rmI}$
the 1-loop self-energies
for those modes.\footnote{In this work $\mPi_{\rmL}$ denotes
the minus of ``$\mPi_{\rmL}$'' in refs. \cite{ta3,ta4}.}
$a_{\rmI}^{n}$ give the contributions involving $n$ bubbles
for the $\rmI$-mode,
such that $A_{\rme \rmI}= a_{\rmI}^{0}$
and $A_{\rmc \rmI} = \sum_{n \ge 1} (-1)^{n - 1} a_{\rmI}^{n}$.

If the integrand of eq. \refer{eq:vfi} given by eqs. \refer{eq:fs},
\refer{eq:yu}
were finite everywhere in the integration
region, the result would coincide with the naive dimension counting
which leads to $\delta v_{\rmF \rmx} \rightarrow {\rm const}$ as
$p_{\rm F} \rightarrow \infty$.
For $\pF \rightarrow \infty$, however,
$\kappa_{\rmI}^{2} \rightarrow 0$
and therefore the integral diverges at $\theta = 0$.
This divergence implies that $\delta v_{\rmF \rmx}$
could contain positive powers or logarithms of $\pF$.
(It can be shown \cite{ta5} that
there is no other singularity.)
$\kappa_{\rmI}^{2} \rightarrow 0$ corresponds formally to
zero meson masses; therefore, these singularities
for $\theta \simeq 0$,
which now prove to give the most dominant contributions
for $\pF \rightarrow \infty$,
correspond to the infrared singularity in the massless limit.

By substituting the analytic formulae of $h_{\rmI}$ \cite{ta4}
and $\mPi_{\rmI}$ \cite{ta1,ku} into eqs. \refer{eq:vfi}-\refer{eq:yu}
and by extracting the singularities for $\theta \simeq 0$,
we can obtain $\delta v_{\rmF \rmx}$
up to logarithmic accuracy.
For $\rmx = \rme$, we obtain
\begin{eqnarray}
\delta v_{\rmF \rme} &=& \left(g_{\si}^{2}\times O(1)\right)
+ \left(\frac{1}{2}\left(\frac{g_{\om}}{2 \pi}\right)^{2}
\left\{\ln\frac{p_{\rmF}^{2}}{m_{\om}^{2}} + O(1) \right\}\right)
\nonumber \\
&+& \left(- \frac{1}{2} \left(\frac{g_{\om}}{2 \pi}\right)^{2}
\left\{\ln\frac{p_{\rmF}^{2}}{m_{\om}^{2}} + O(1) \right\}\right).
\label{eq:d1}
\end{eqnarray}
The three terms show the contributions due to the
relevant three modes $\si$, $\rmL$ and $\rmT$, respectively.
Though both the contributions due to
the L- and T-modes grow logarithmically, reflecting
the logarithmic infrared divergence at $\theta \rightarrow 0$,
they cancel
out, leading to $\delta v_{\rmF \rme} = {\rm const}$.
The contribution to $\delta v_{\rmF \rmc}$ due to one
bubble insertion,
i.e., due to the $n = 1$ term of
eq. \refer{eq:yu},
can be obtained similarly:
\begin{eqnarray}
\delta v_{\rmF 1 {\rm b}} &=&
\left(g_{\si}^{4} \times O(\ln p_{\rmF}^{2})\right)
+ \left(
- \left( \frac{g_{\om}}{2 \pi}\right)^{4}
\left\{ \frac{4 p_{\rmF}^{2}}{m_{\om}^{2}} - \frac{1}{3}
\ln \frac{p_{\rmF}^{2}}{M^{2}}
\ln\frac{p_{\rmF}^{2}}{m_{\om}^{2}}+ O(\ln p_{\rmF}^{2})
\right\}\right)
\nonumber \\
&+& \left(- \left(\frac{g_{\om}}{2 \pi}\right)^{4}
\left\{ \frac{1}{3} \ln \frac{p_{\rmF}^{2}}{M^{2}}
\ln \frac{p_{\rmF}^{2}}{m_{\om}^{2}} + O(\ln p_{\rmF}^{2})
\right\}
\right).
\label{eq:d2}
\end{eqnarray}
The coupling to the ph excitations makes the infrared
divergences stronger;\footnote{``h'' includes the
negative-energy states.}
the most dominant contribution is due to
the linearly divergent integral
due to the L-mode,
and the result grows like $p_{\rmF}^{2}$.
However, it can be easily seen that the insertion of
more bubbles causes stronger infrared divergences,
leading to terms of higher powers in $\pF$.

This situation is reminiscent of the high-density
NR electron gas \cite{ge}.
This suggests that we have to sum up all
the bubble graphs, forming the complete RPA series.
Therefore, {\it the infrared structure of $v_{\rmF}$
due to the coupling of the ph excitations
naturally leads to our $1/N$ expansion scheme.}
By summing up all the
bubbles,
we obtain for the total $\delta v_{\rmF}$:
\begin{eqnarray}
\delta v_{\rmF} &=& \left(g_{\si}(p_{\rmF})^{2}\times O(1)\right)
+ \left(g_{\om}(p_{\rmF})^{2} \times O(1) \right)
\nonumber \\
&+& \left(- \frac{1}{2} \left(\frac{g_{\om}(p_{\rmF})}{2 \pi}\right)^{2}
\left\{\ln\frac{p_{\rmF}^{2}}{m_{\om}^{2}} + O(1) \right\}\right).
\label{eq:d3}
\end{eqnarray}
The resummation of the ring graphs renders the infrared
divergence mild, leaving only the logarithmically growing term
due to the $\rmT$-mode (compare the dashed and the solid curves
of fig. \ref{fig:fig4}).
Thus the decrease of $v_{\rmF}$, which led
to the inversion of the spectrum of fig. \ref{fig:fig5},
is due to the logarithmic correction \refer{eq:d3}
due to the self-energies
beyond the mean-field theory.

The resummation
also replaces the original coupling constants $g_{\rmI}$
renormalized at the meson momenta $q^{2} = 0$
by the running coupling constants $g_{\rmI}(\pF)$
at the scale $\pF$ \cite{ta5}:
Before applying Redmond's method
to avoid the Landau ghost \cite{re}, they are related
by the one-loop formulae:
for $\pF \gg M$,
$g_{\om}(\pF)^{2} \simeq g_{\om}^{2} \left/
\left(1 - \frac{g_{\om}^{2}}
{6 \pi^{2}} \ln\frac{p_{\rmF}^{2}}{M^{2}} \right) \right.,$
and similarly for $g_{\si}(\pF)$.
$g_{\rmI}(\pF)$ of eq. \refer{eq:d3} should be understood
as the result of Redmond's method;
they grow for increasing $\pF$
toward finite values for the bare coupling constants
\cite{re}.
Thus the growth of the last term of eq. \refer{eq:d3}
is enhanced by the growth of $g_{\om}(p_{\rmF})$.

Our result can be summarized
by the behavior of the screening mass in the medium.
The infrared behavior of the $\rmI$-mode contribution
is governed by the screening mass
(squared) $m_{\rmsc \rmI}^{2}$,
which is obtained from the self-energy $\mPi_{\rmI}(q)$
by letting
$q^{0} \rightarrow 0$ first followed by $\bfq \rightarrow 0$
(see eqs. \refer{eq:vfi}-\refer{eq:yu}).
The Lorentz invariance
guarantees $m_{\rmsc \rmL}^{2} = m_{\rmsc \rmT}^{2}$,
which would lead to the cancellation between the $\rmL$- and $\rmT$-mode
contributions (see eq. \refer{eq:d1}).
However, the coupling of the ph excitations
breaks the invariance:
The $\rmL$-mode acquires a nonzero Debye screening mass,
which is
$m_{\rmsc \rmL}^{2} = 2 g_{\om}^{2} p_{\rmF}^{2}/\pi^{2}$
for $\pF \rightarrow \infty$, but
$m_{\rmsc \rmT}^{2} = 0$
due to baryon current conservation \cite{be}.
As a result, the cancellation between the two modes becomes
incomplete, leaving the result of eq. \refer{eq:d3}.

Now we discuss physical interpretations of our results.

Above the critical $p_{\rmF}^{\rmc}$ where $v_{\rm F} = 0$,
the normal state is unstable.
In our calculation of the ground-state
energy density and
the quasiparticle energy
we assumed the normal state \cite{la1}.
If $v_{\rmF} < 0$, however,
this assumption is not valid.
This suggests
a stability condition: $\vF > 0$ for a normal Fermi liquid,
in addition to the wellknown conditions
for the dimensionless Landau-Migdal parameters \cite{la1}.
If the ground state is not normal,
one should re-compute the energy density self-consistently
allowing the possibility of new configurations
like the so-called ``Fermi-gap'' state:
If one literally accepts the results of fig. \ref{fig:fig5},
this indicates a 1st order phase transition
at $\pF = p_{\rmF}^{\rmc}$
from the normal state to the Fermi-gap state.\footnote{The possibility of
the Fermi-gap state for high-density nuclear matter
was pointed out in ref. \cite{ho}
in the Hartree-Fock approximation of the lagrangian \refer{eq:la}.
However, as stressed above, the mechanism in our case is
beyond the Hartree-Fock approximation (recall
$\delta v_{\rmF \rme} = {\rm const}$).}
In contrast to the Fermi sphere
for the normal state, the nucleons occupy the momentum states
within the inner sphere
and the outer shell
for the Fermi-gap state.
In our model, $p_{\rmF}^{\rmc} = 4.07$fm$^{-1}$ which would
imply the critical density for neutron matter
$\varrho^{\rmc} \gaq 15\ro_{0}$.

We also note another possible scenario:
Near the critical point $v_{\rm F} = 0$,
the terms of higher order than we considered here,
which would give contributions
$\sim ( g_{\om}(p_{\rmF})^{2} \ln p_{\rmF}^{2} )^{n}$ ($n \ge 2$),
also could be important.
It would be a very hard job to compute these contributions.
Intuitively, one could regard
the result of
eq. \refer{eq:d3} as the leading approximation
of $\vF \approx 1/
\left(1 + \left( g_{\om}(\pF)/2 \pi\right)^{2}/2
\times \ln \left(p_{\rmF}^{2}/m_{\om}^{2}\right) \right)$.
In this case $\vF = 0$ would not occur for any finite $\pF$;
$\vF \rightarrow 0$ asymptotically.

In connection with these scenarios,
we stress that the (formal) high-density limit
based on the lagrangian \refer{eq:la} does not correspond to the
mean-field theory, contrary to the usual claim \cite{se}:
For the first scenario this is obvious; the second one
implies an infinite density of states at the Fermi surface
for $\pF \rightarrow \infty$,
which never be reached in the mean-field theory.
The usual claim tacitly assumes the absence of infrared
singularities, which reflect large long-wavelength
fluctuations of the fields,
from the higher order corrections.
This assumption leads to the validity of the naive dimension counting,
and therefore to the dominance of the mean-field contributions
because they contain the largest number of independent fermion loops
in each order of the interaction.
As discussed above, the effects which lead to the modification
of this naive argument are
due to the strong infrared behavior
induced in the high-density medium.

This conclusion for the high-density behavior
seems to be inevitable for relativistic descriptions:
The relevant infrared behavior is due to the ph excitation in the
vector channel (see eqs. \refer{eq:d1}-\refer{eq:d3}).\footnote{In
view of this, the extension of our approach
to gauge theories would be appealing,
and is under investigation
for cases
of cold QED and QCD plasma with finite fermion density \cite{ta5}.}
However, the vector boson
degrees of freedom are indispensable to keep
matter from collapsing at high density by supplying the repulsion.
We also note that the conclusion might be valid
even if the compositeness of hadrons were taken into account:
Even though the vertex form factors were introduced,
they could not influence the
results dominated by the contribution from
the infrared region ($q^{2} = 0$).

Though we have shown the possibility
of the Fermi-gap state, the transition density $\varrho^{\rmc}$
appears to be above the critical densities
to more familiar new phases like
quark matter phase, pion condensed phase, etc.
However, our result was obtained just within a simple model,
and $\varrho^{\rmc}$
is very sensitive even to small corrections,
which might be due to higher orders in $1/N$
or other hadronic
degrees of freedom like $\pi$, $\rho$, etc.
After the inclusion of these effects,
the competition or coexistence of the Fermi-gap
state with more familiar new phases
would be an interesting aspect of high-density
hadronic matter.

In conclusion, we have investigated dynamical effects
in high-density nuclear matter
due to the higher order many-body correlations.
The anomaly in the fermion spectra,
corresponding to the famous effective mass enhancement in NR
Fermi liquids, grows with increasing density.
The effect is due to the strong infrared behavior caused by
the coupling of the ph excitation to the single-particle motion.
The picture of high-density matter emerged from our investigation
is much more complicated than the mean-field description.
Though a more realistic treatment
would be required to fix the relevance
to phenomenology,
the universal character of this phenomenon already poses various
interesting aspects for high-density matter,
like the possibility of the Fermi-gap state, the thermodynamic
properties and transport phenomena in this phase, and
the extension to gauge theories.
These points will be reported in future publications \cite{ta5}.

\vspace{0.5cm}
The author expresses his thank to Dr. W. Bentz for helpful discussions.
He is also grateful to Prof. K. Yazaki and the members of nuclear
theory group in RIKEN for valuable comments.
The work was performed under the auspices of Special Researchers'
Basic Science Program of RIKEN.

\newpage

\section*{Figure captions}
\begin{enumerate}

\item{Feynman graphs of the nucleon self-energy of $O(1/N)$.
The dashed lines in the figure denote the noninteracting
$\si$- and $\om$-meson
propagators;
the solid lines denote the nucleon Hartree
propagators on the background meson fields
and the Fermi sea.}
\label{fig:fig2}

\item{The Fermi velocity as a function of the Fermi momentum.
For explanation of the curves, see text.}
\label{fig:fig4}

\item{The real parts of the quasiparticle energy (eq. \refer{eq:qp2})
are shown as functions of
momentum by the solid lines. (a), (b) and (c) are
for cases of $p_{\rmF}=1.30$, $4.07$ and $4.50{\rm fm}^{-1}$,
which respectively corresponds to $\vF > 0$, $= 0$ and $<0$.
The dot-dashed line shows the Hartree contribution contained
in the full result.
The insertion in (b) shows
the $O(1/N)$-contributions as functions of momentum.
The dot-dashed line shows $\epsilon_{\rme}$ of eq. \refer{eq:qp2},
while the solid line shows $\epsilon_{\rmc}$.
The dotted and dashed lines are the separated contributions
$\epsilon_{\rmc \rmW}$ and $\epsilon_{\rmc \rmP}$
contained in the solid line.}
\label{fig:fig5}

\end{enumerate}

\section*{Table captions}
\begin{enumerate}

\item{The parameters used for numerical computation.}
\label{tab:tabn}

\end{enumerate}

\newpage

\newcommand{\mev}{{[MeV]}}
\large

\section*{Table 1}
\begin{tabular}{|l||rrcc|} \hline
     & $\displaystyle{g_{\sigma}^{2}}/4 \pi$&
$\displaystyle{	g_{\omega}^{2}}/4 \pi$&
$m_{\sigma}$ \mev &
$m_{\omega}$ \mev \\ \hline
Hartree& 6.23 & 8.18 & 550 & 783\\
      &      &      &     &    \\
$1/N$  \hspace{0.42cm}A& 2.24 & 3.65 & 550 & 783\\
\quad \qquad B      & 2.60 & 4.22 & 650 & 783\\
      &      &      &     &      \\
\hline
\end{tabular}

\end{document}